\documentclass[aps,prd,nofootinbib,superscriptaddress,preprint,amssymb,12 pt]{revtex4-1}
\usepackage{cancel}
\usepackage{amsfonts}
\usepackage{amsmath}
\usepackage{booktabs}
\usepackage{multirow}
\usepackage{slashed}
\usepackage{graphicx,color}
\usepackage{subfigure}
\usepackage{xcolor}
\usepackage{enumerate}
\usepackage[
            citecolor=blue,
            pdfstartview=FitH,
            bookmarksnumbered=true,
            bookmarksopen=true,
            anchorcolor=blue,
            menucolor=red,
            linkcolor=blue,
            filecolor=red,
            runcolor=red,
            urlcolor=blue,
            frenchlinks=red
            ]{hyperref}
\usepackage{indentfirst}%LaTeX

\hfuzz=\maxdimen
\begin{document}

\title{ Precise evaluation of $h\to c\bar{c}$ and  axion-like particle production}

\author{Shi-Yuan Li}\email{lishy@sdu.edu.cn}
\affiliation{School of Physics, Shandong University, Jinan, Shandong 250100,  China}
\author{Zhen-Yang Li}\email{201812125@mail.sdu.edu.cn}
\affiliation{School of Physics, Shandong University, Jinan, Shandong 250100,  China}
\author{Peng-Cheng Lu}\email{pclu@sdu.edu.cn}
\affiliation{School of Physics, Shandong University, Jinan, Shandong 250100,  China}
\author{Zong-Guo Si}\email{zgsi@sdu.edu.cn}
\affiliation{School of Physics, Shandong University, Jinan, Shandong 250100,  China}

\begin{abstract}
We study the decay of  the SM Higgs boson to a massive charm quark pair at the next-to-next-to-leading order  QCD  and next-to-leading order  electroweak.
At the second order of QCD coupling,  we consider the exact calculation of flavour-singlet contributions where the Higgs boson couples to
the internal top and bottom quark.  Helpful information on the running mass effects related to Yukawa coupling may be obtained by analyzing this process.
High precision production for $h\to c\bar{c}$ within the SM makes it possible to search for new physics that may induce relatively large interactions related to the charm quark. As an example, we evaluate the  axion-like particle  associate production with a charm quark pair in the Higgs decay and obtain some constraints  for the corresponding parameters under some assumptions.
\end{abstract}

\maketitle

%++++++++++++++++++++++++++++++++++++++++++++++++++++++++
\section{Introduction}

Higgs physics plays a significant role in the standard model(SM) testing and search for new physics beyond SM with higher precision. The theoretical and experimental study on
the properties of the Higgs boson, including its couplings with SM particles, particularly fermions, has become a primary task of particle physics at the high energy frontier
 since it was discovered in 2012 at the LHC~\cite{Aad:2012tfa,Chatrchyan:2012ufa}, and considerable efforts have been made.   The study of the process of decay of the Higgs boson to heavy quark
 pairs $b\bar b$ and $c\bar c$  is important to test the SM and search for high precision new physics.
Furthermore, useful information for the running mass effects in the Yukawa coupling can be obtained by analysing these decay processes.
 The decay width of Higgs boson into massless bottom quarks is known up to the
next-to-next-to-next-to-next-to-leading order in QCD~\cite{Baikov:2005rw,Davies:2017xsp,Herzog:2017dtz}.
 The differential decay width for $h\to b \bar{b}$ has been computed to the next-to-next-to-leading order (NNLO) QCD
 in \cite{Anastasiou:2011qx,DelDuca:2015zqa}, and next-to-next-to-next-to-leading order in \cite{Mondini:2019gid}
 in the limit where the mass of the bottom quark is neglected. The corrections keeping the
exact bottom quark mass  have been computed to
NNLO in \cite{Bernreuther:2018ynm,Caola:2019pfz,Behring:2019oci,Somogyi:2020mmk}. The hadronic decays of Higgs boson to the bottom quarks, light quarks, and gluons at NNLO have been calculated in \cite{Hu:2021rkt}.
For the two loop QCD corrections of $h\to b \bar{b}$,  the flavor-singlet contribution from the triangle top and bottom quark loop have to be  included.
These contributions from the top quark triangle loop  are computed approximately using a large quark mass expansions formula \cite{Larin:1995sq}, and  the corresponding exact calculations
 are also obtained \cite{Primo:2018zby}. The calculations at the next-to-leading order(NLO) electroweak(EW) are also completed \cite{Dabelstein:1991ky,Fleischer:1980ub,Kniehl:1991ze}.
The measurements of $h\to b \bar b$  at LHC are finished \cite{Sirunyan:2018kst,Aaboud:2018zhk}, and the signal strength $\mu_{h\to b \bar b} = 1.01\pm 0.12 ~(\mathrm{stat.}) ^{+0.16}_{-0.15} ~(\mathrm{syst.})$ is compatible with SM.

For the evaluation of Higgs boson coupling to the second generation quark, $h\to c\bar{c}$ is a process of great value.
Owing to the absence of an observation of Higgs decays to the charm quark, there are some
bounds on the charm quark Yukawa coupling. An indirect search for the decay of the Higgs boson to charm quarks via the decay to $J/\psi \gamma$ is presented
 at the LHC \cite{Perez:2015aoa,Aad:2015sda,Sirunyan:2018fmm}. A direct search for the Higgs boson, produced in association with a vector boson ($W$ or $Z$),
and decaying to a charm quark pair is performed by CMS collaboration \cite{Sirunyan:2019qia} and ATLAS collaboration \cite{Aaboud:2018fhh}, respectively.
 This result is the most stringent limit to date for the inclusive decay of the $h\to c\bar{c}$. Both ATLAS and CMS collaborations present novel methods for charm-tagging.  To test the SM  and search for new physics in high precision,
the theoretical predictions and experimental measurements on SM Higgs couplings should be completed as precisely as possible. To achieve this,
we finish the exact calculation for $h\to c \bar c$ up to NNLO QCD including the flavor-singlet contributions from the triangular loops of the bottom and top quarks and those at the NLO EW.
However, owing to the discovery of the Higgs boson, the search for additional (pseudo) scalar bosons  beyond the SM has attracted increasing interest in collider physics.  The  axion-like particles (ALPs) are a hypothetical (pseudo)scalar  that  naturally arises
in many extensions of the SM as pseudo Nambu-Goldston bosons.  The ALP parameter space has been intensively explored \cite{Cadamuro:2011fd,Millea:2015qra,DiLuzio:2016sbl}, covering a wide energy range \cite{Mimasu:2014nea,Bauer:2017ris,Brivio:2017ije,Alonso-Alvarez:2018irt,Harland-Lang:2019zur,Baldenegro:2018hng}. These experimental searches allow access to several orders of magnitude in the ALP masses and couplings \cite{Irastorza:2018dyq}, where astrophysics and cosmology impose constraints in the sub-KeV mass range and the most efficient probes in the MeV-GeV range are obtained from experiments acting on the precision frontier \cite{Essig:2013lka}.
Another aim of this study is to evaluate the ALPs that dominantly couple to the up type quark \cite{Carmona:2021seb}, e.g., charm quark, and to obtain some constraints on the corresponding
parameters from the ALP associate production with the charm quark pair in the Higgs decay.

This paper is organized as follows. In Sec.\ref{Sec:h2cc}, we study the decay width of $h\to c\bar{c}$ within the SM. First, we use renormalized matrix elements where the QCD coupling is defined in the
$\overline{\text{MS}}$ scheme, whereas  the charm quark mass and the Yukawa coupling are defined in the on-shell scheme. For the order of $\alpha_s^2$, we calculate the exact flavor-singlet contributions. To obtain reliable corrections to the decay width of $h\to c\bar{c}$, we express the on-shell Yukawa coupling in terms of the $\overline{\text{MS}}$ Yukawa coupling.
At the end of this section, after including the NLO EW corrections,  we obtain the decay width of $h\to c\bar{c}$ and compare it with that of $h\to b\bar{b}$.
In Sec.\ref{Sec:ALP}, we evaluate the ALP associate production with the charm quark pair in Higgs decay.  Finally, a brief summary is given.

\section{Precise evaluation  of the $h\to c\bar{c}$ process}\label{Sec:h2cc}

Within the SM, the Higgs interactions with quarks are obtained by considering the Yukawa interactions
\begin{equation}
-{\cal L}_Y\, =\, \overline{U_R} h_u Q^T (i\tau_2)\Phi - \overline{D_R} h_d Q^T (i\tau_2)\tilde{\Phi}+h.c.,
\end{equation}
where $\Phi$ denotes a complex $SU(2)_L$ doublet Higgs field with hyper charge $Y=1/2$.
$\tau$ is the usual $2\times 2$ Pauli matrix; $\tilde{\Phi}=i\tau_2\Phi^*$.
$Q^T=(U_L,D_L)$ where $U$ and $D$ represent three up- and down-type quarks; and $h_{u,d}$ represents Yukawa matrices. By choosing $\Phi=(0,v+h)^T/\sqrt{2}$, we obtain
\begin{equation}
-{\cal L}_{hf\bar f}\, =\, y_{0,f} h f \bar{f},~~~ f=u,d,s,c,b,t,
\end{equation}
where $y_{0,f}=m_{0,f}/v$ is the bare Yukawa coupling constant  and $m_{0,f}$ is
 the bare mass of the corresponding quark.
The Higgs vacuum expectation value
$v=(\sqrt{2}G_F)^{-1/2}$ with $G_F$  the Fermi constant.

In our following  analysis,
we start by using the  renormalized matrix elements where the QCD coupling constant $\alpha_s$ is defined in the $\overline{\text{MS}}$ scheme, whereas the charm
quark mass  $m_c$ and the related
 Yukawa coupling constant $y_c$  are defined in the on-shell scheme.
 Up to NNLO QCD, the decay width of the process $h\to c\bar{c}$ can be written as follows
 \begin{align}
\Gamma\, =\, \Gamma_{LO}+\alpha_s(\mu)  \tilde{\Gamma}_1+\alpha_s^2(\mu) \tilde{\Gamma}_2+
\frac{\alpha_s^2(\mu)}{2\pi} \,
 \ln \Big(\frac{\mu^2}{m_c^2} \Big)\, \Big[\frac{11}{6}N_c-\frac{n_f+1}{3}\Big]  \tilde{\Gamma}_1~,
\label{equ:Gamma1}
 \end{align}
where $\mu$ is the renormalization scale.  $N_c$ and $n_f$ denote
the number of the color and  the number of the massless quarks, respectively.
The decay width at the leading order(LO) can be expressed as follows
\begin{align}
  \Gamma_{LO}\, =\, N_C~\frac{y_c^2}{8\pi m_h^2}(m_h^2-4m_c^2)^{3/2}~,
\label{equ:LO}
\end{align}
where $m_h$ and $m_c$ denote the on-shell  mass of the Higgs boson and charm quark, respectively. $\tilde{\Gamma}_1$ and $\tilde{\Gamma}_2$ respectively  represent the
scaled next-to-leading order(NLO) and NNLO decay width of $h\to c\bar{c}$ by the factor out $\alpha_s$.
At the NLO QCD, the scaled decay width $ \tilde{\Gamma}_1$  can be expressed as follows \cite{Drees:1990dq}
\begin{align}
\tilde{\Gamma}_1\, =\, &\Gamma_{LO} \, C_F\, \delta_1 ~,
\label{equ:NLO}
 \end{align}
where  $C_F=(N_c^2-1)/2/N_c$, and
\begin{align}
\delta_1\, =\, &\frac{1}{\pi} \,    \,\Big\{
\frac{A(\beta)}{\beta}\,-\, \frac{3+34\beta^2-13\beta^4}{16\beta^3}\ln (x)\,  -\, \frac{3(1-7\beta^2)}{8\beta^2}
 \Big\},
\label{equ:NLO}
 \end{align}
 with   $\beta=\sqrt{1-4m_c^2/m_h^2}$, $x=(1-\beta)/(1+\beta)$ and
\begin{align}
A(\beta) \, =\, & (1+\beta^2)\Big\{
4Li_2(x)+2Li_2(-x)\, +\,  2\ln(\beta)\ln(x)\,+\, 3\ln \Big(\frac{1}{1+\beta} \Big)\ln(x)\Big\}\nonumber \\
& -3\beta \ln \Big(\frac{4}{1-\beta^2} \Big)-4\beta \ln(\beta).
 \end{align}
At the NLO QCD, the large logarithm related to the ratio of the charm-Higgs mass appears. This effect can be clearly shown in the limit $m_c\to 0$, i.e.,
\begin{align}
\delta_1\, \simeq \, &\frac{3}{2\pi} \,\Big( \,  \frac{3}{2}-\ln\frac{m_h^2}{m_c^2} \, \Big).
 \end{align}
The  large logarithm $\ln (m_h^2/m_c^2)$ should be absorbed in the running charm
quark mass in the Yukawa coupling as discussed below.
$\Gamma_{LO}$, $\tilde{\Gamma}_1$
and $\tilde{\Gamma}_2$ do not depend on $\mu$. The $\mu$ dependence of the last term in eq.(\ref{equ:Gamma1}) can be obtained from the renormalization group equation.

 \begin{figure}[!ht]
   \centering
   \includegraphics[scale=0.8]{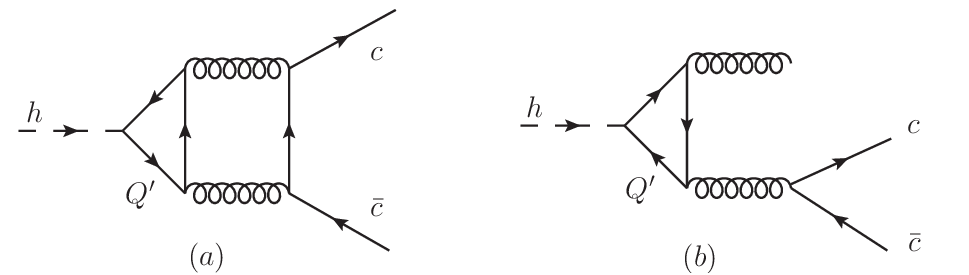}
   \caption{The flavor-singlet $\mathcal{O}(\alpha_s^2)$ contributions to $h \to c \bar{c}$(a)  and $h \to c \bar{c}g$(b) with  $Q' = t, b, c$.}
   \label{fig:SQ}
\end{figure}

At the NNLO QCD, the subtraction term of double-real radiation corrections and real-virtual corrections were computed in \cite{Dekkers:2014hna,Chen:2016zbz,Bernreuther:2011jt,Bernreuther:2013uma}. For the double-virtual corrections,
 the scaled decay width $\tilde{\Gamma}_2$ receives the flavor-singlet and non-singlet contribution.  The
flavour-singlet contributions from $h\to c\bar{c}$  (fig.\ref{fig:SQ}a) and $h\to c\bar c g$( fig.\ref{fig:SQ}b) are both UV and IR finite separately.
Because we keep the non-zero internal quark mass $m_f$($f=c, b, t$),   these three kinds of massive quarks, which couple to the Higgs in the triangle loop,  contribute
to the decay width of $h\to c\bar{c}$.  For the calculation  of the case  where the Higgs couples to the internal charm quark, i.e., equal masses associated
with  the  inner  and  outer  fermion  line, and that of
the flavor non-singlet contribution,  the strategy and formulas can be found in \cite{Bernreuther:2018ynm,Bernreuther:2005gw,
Bernreuther:2005rw,Bernreuther:2004ih,Bernreuther:2004th} .  Here, we focus on discussing the flavor-singlet contribution to
$h\to c\bar{c}$ where the Higgs boson couples to the bottom or top quark.  For fig.\ref{fig:SQ}a with $Q'=b$ or $t$, we calculate the exact results  using  the formula and techniques of \cite{Primo:2018zby} where  the Mathematica packages $\mathrm{PolyLogTools}$ \cite{Duhr:2019tlz},  $\mathrm{HPL}$, and $\mathrm{GINAC}$ library are necessary. For fig.\ref{fig:SQ}b,
we finish the calculation independently. In our numerical analysis, we choose $m_h=125.09$ GeV,
$v\simeq 246.2$ GeV,
$m_c=1.68$ GeV, $m_b=4.78$ GeV,  $m_t=173.34$ GeV, $\alpha_s(m_Z)=0.118$, and $m_Z=91.2$ GeV \cite{Tanabashi:2018oca}.
 In table \ref{tab:os1}, we list our results   $\tilde{\Gamma}_{Q'}^{FS}$ for the flavor-singlet contribution with $Q'=b$, $t$ where $\alpha_s^2$ is factored out in the partial decay width.
It is found that $\tilde{\Gamma}_{b}^{FS}$ is considerably smaller than $\tilde{\Gamma}_{t}^{FS}$.
For the case where top quark appears in the triangle loop,   $m_c\leqslant m_h \leqslant m_t$, and  $\tilde{\Gamma}_{t}^{FS}$ can be calculated
to the leading order in the charm quark mass as an expansion in the inverse powers of the top quark mass \cite{Larin:1995sq}. It is found that under this approximation,
$\tilde{\Gamma}_{t}^{FS}\simeq 0.7417$ MeV which is very close to the exact result. For the triangular loop related to the bottom quark, this approximation is  unreliable.
At the NNLO QCD, the other two interesting sub-processes are $h\to c\bar{c} c\bar{c}$ and
$h\to b \bar{b} c\bar{c}$. The results for the scaled decay width $\tilde{\Gamma}_{c\bar{c} c\bar{c}}$ and $\tilde{\Gamma}_{b\bar{b} c\bar{c}}$ where $\alpha_s^2$ is also factored out
are listed in table I. Obviously, $\tilde{\Gamma}_{c\bar{c} c\bar{c}}$ is significantly smaller  than $\tilde{\Gamma}_{b\bar{b} c\bar{c}}$, but for  $\tilde{\Gamma}_{b\bar{b} c\bar{c}}$,
the contribution from the interaction between the Higgs boson and bottom quark is dominant. As a result, the contribution from  $h\to b \bar{b} c\bar{c}$ should not be included in the decay
width of $h\to c\bar{c}$.
Up to the NNLO QCD, the ratio of the  flavor-singlet contribution with $Q'=b$, $t$ to the total decay width is approximately  $4\% $.
\begin{table}[!ht]
  \centering \caption{Results for the scaled decay width $\tilde{\Gamma}_{b}^{FS}$, $ \tilde{\Gamma}_{t}^{FS}$,  $\tilde{\Gamma}_{c\bar{c} c\bar{c}}$  and $\tilde{\Gamma}_{b\bar{b} c\bar{c}}$.}\label{tab:os1}
\begin{tabular}{|c|c|c|c|}\hline\hline
~~$\tilde{\Gamma}_{b}^{FS}$~[MeV]~~  &~~$\tilde{\Gamma}_{t}^{FS}$~[MeV]~~   & ~~$\tilde{\Gamma}_{c\bar{c} c\bar{c}}$~[MeV]~~  & ~~$\tilde{\Gamma}_{b\bar{b} c\bar{c}}$~[MeV]~~\\[1pt]\hline
~~~-0.0482~~~      & ~~~0.7411~~~   &  ~~~0.4262~~~   &~~~3.4495~~~   \\\hline \hline
\end{tabular}
\end{table}

It is known that under the on-shell renormalization scheme for the Yukawa coupling between the SM Higgs and massive quarks, the large logarithm of the fermion-Higgs boson mass ratio can be obtained in the radiative corrections \cite{Braaten:1980yq}. Therefore, it is not a good idea to study the
decay width in terms of the on-shell Yukawa coupling when $m_c / m_h \ll 1$. When using the $\overline{MS}$ Yukawa coupling
\begin{equation}
\overline{y}_c(\mu)\, =\, \overline{m}_c(\mu)/v~,
\end{equation}
where $\overline{m}_c(\mu)$ is the running $\overline{MS}$ mass and $\mu=m_h$;
this kind of large  logarithmic effects can be reduced
by effectively absorbing all relevant large logarithms into the running mass related to the Yukawa coupling constant. In this study,  to obtain  reliable
corrections to the decay width of $h\to c\bar c$,  we converted Yukawa coupling constant $y_c$ defined in the on-shell scheme to $\overline{y}_c$ in the
$\overline{\text{MS}}$ scheme.
The relation between on-shell mass $m_c$ and $\overline{\text{MS}}$ mass $\overline{m}_c(\mu)$ can be written as follows \cite{Gray:1990yh}:
 \begin{align}
  m_c\, =\, \overline{m}_c(\mu)~\Big[1+c_1(m_c,\mu)\frac{\alpha_s(\mu)}{\pi}
  +c_2(m_c,\mu) \Big(\frac{\alpha_s(\mu)}{\pi} \Big)^2 \Big]+\mathcal{O}(\alpha_s^3)~,
\label{equ:mQ}
 \end{align}
where
\begin{align}
c_1\, =\, & C_F(1+\frac{3}{4}L_c),\nonumber \\
c_2\, =\, &C_F^2\, [  \frac{121}{128}+3\zeta(2)(\frac{5}{8} - \ln2)+\frac{3}{4}\zeta(3)
+\frac{27}{32}L_c+\frac{9}{32}L_c^2]
\nonumber  \\
     &-N_C C_F \, [-\frac{1111}{384}+\frac{\zeta(2)}{2}(1-3\ln2)+\frac{3}{8}\zeta(3)-\frac{185}{96}L_c-\frac{11}{32}L_c^2]  \nonumber \\
     &-C_F T_F n_f \, [\frac{71}{96}+\frac{1}{2}\zeta(2)+\frac{13}{24}L_c+\frac{1}{8}L_c^2]  \nonumber \\
    &-C_F T_F \, [\frac{143}{96}-\zeta(2)+\frac{13}{24}L_c+\frac{1}{8}L_c^2],
\label{equ:u1u2}
\end{align}
%%%%%%%%%%%%%%%%%%%%%%%%%%%
with $T_F=1/2$, $L_c=\ln({\mu^2}/{m_c^2})$,  $\zeta(2)={\pi^2}/{6}$  and $\zeta(3)=1.20205690...$.

\begin{figure}[!ht]
   \centering
   \includegraphics[scale=0.5]{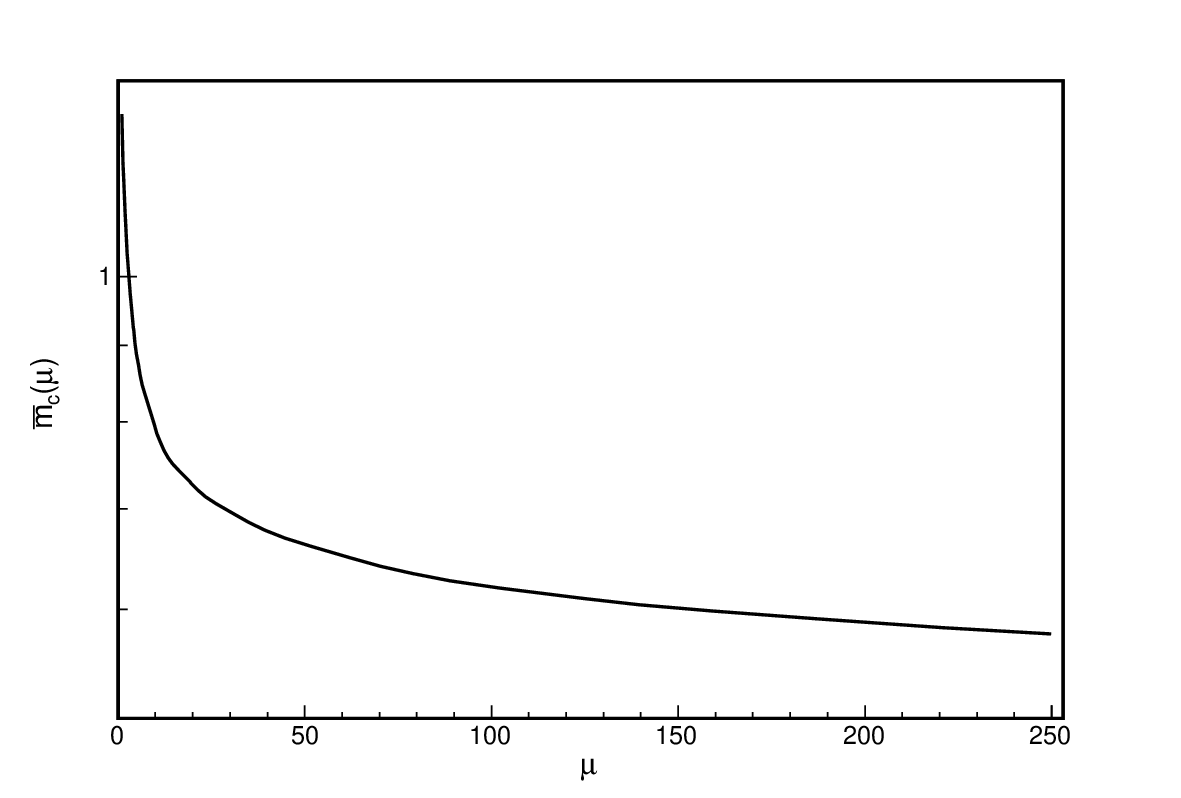}
   \caption{Dependence of $\overline{m}_c(\mu)$ on $\mu$.}
   \label{fig:mcbar}
\end{figure}

To compute the  $\overline{\text{MS}}$ Yukawa coupling $\overline{y}_c(\mu)$ for the arbitrary scale $\mu$,
we use the solution of the renormalization group equation for $\overline{m}_c(\mu)$ at two-loops
\begin{eqnarray}
 &&\overline{m}_c(\mu)\, =\, \overline{m}_c(\mu_0)\Big(\frac{\alpha_s(\mu)}{\alpha_s(\mu_0)} \Big)^{1/\beta_0}
\Big \{ 1+\frac{(\gamma_{1}/\beta_0-1/\beta_1)}{\pi\beta_0^2}[\alpha_s(\mu)-\alpha_s(\mu_0) ]+O(\alpha_s^2) \Big\},
 \label{equ:arbitrary m}
\end{eqnarray}
where
\begin{equation}
\gamma_{1}=\frac{303-10n_f}{72}, ~~~ \beta_0=\frac{33-2n_f}{12},~~~\beta_1=\frac{153-19n_f}{24}.
\end{equation}
For the on-shell $c$-quark mass,  we used $\overline{m}_c(\mu=\overline{m}_c)=1.28$~GeV~\cite{Tanabashi:2018oca} as the input in eq.(\ref{equ:mQ})
, and obtained  $m_c=1.68$~GeV.  The results for the dependence of $\overline{m}_c(\mu)$ on $\mu$ are displayed in fig.\ref{fig:mcbar}.

\begin{figure}[ht]
   \centering
   \includegraphics[scale=0.5]{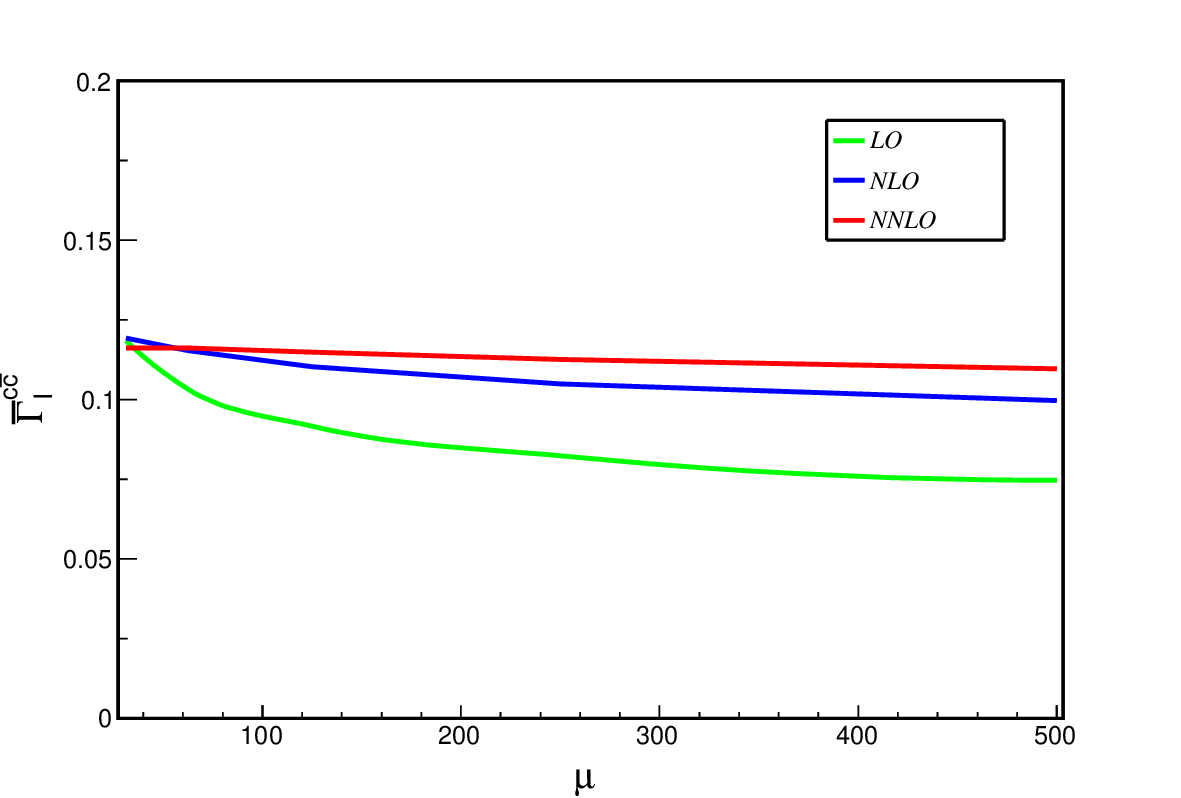}
   \caption{Dependence of $\Gamma_I$ (I=LO,NL0,NNLO) on $\mu$.}
   \label{fig:sta}
\end{figure}

\begin{figure}[!ht]
   \centering
   \includegraphics[scale=0.8]{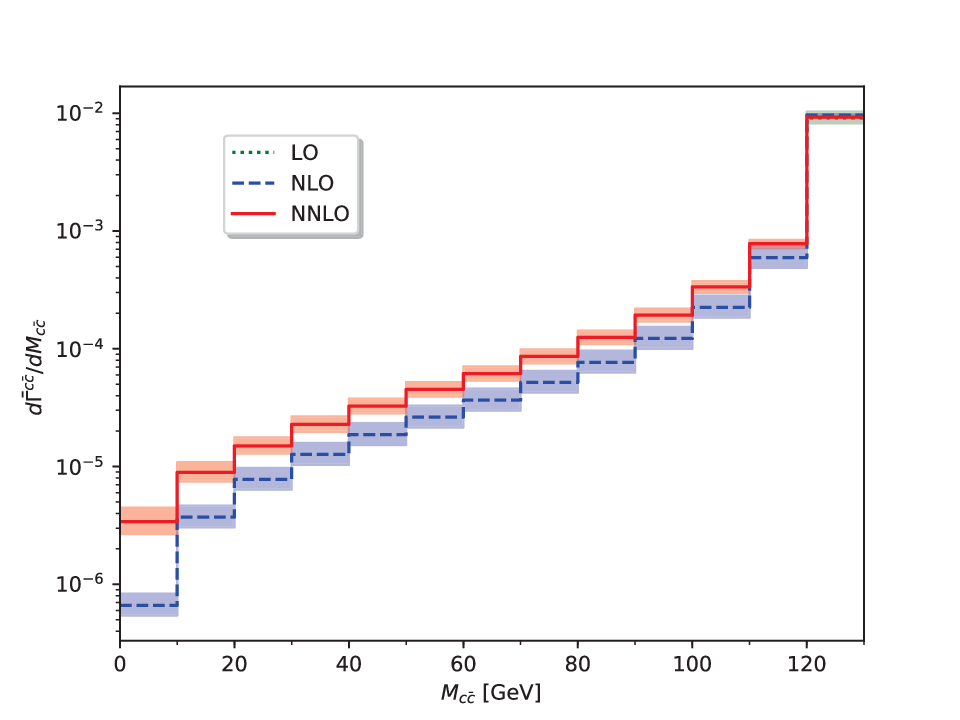}
   \caption{Distribution $d\overline{\Gamma}^{c\bar{c}}/dM_{c\bar{c}}$ of the $c\bar{c}$ invariant mass at LO(short-dashed), NLO(long-dashed), and NNLO(solid) QCD corrections. The short-dashed, long-dashed, and solid lines correspond to the scale choice $\mu=m_h$ whereas the shaded bands show the effect of varying the renormalization scale between $\mu=m_h/2$ and $\mu=2 m_h$.}
   \label{fig:mcc}
\end{figure}

Inserting eq.(\ref{equ:mQ}) into eq.(\ref{equ:Gamma1}), we obtain the decay width for $h\to c\bar c$ calculated  in term of the
$\overline{\text{MS}}$ Yukawa coupling
\begin{eqnarray}
 \overline{\Gamma}&=&\overline{y}_c^2(\mu)\hat{\Gamma}_{0}^{c\bar{c}}[1+\frac{\alpha_s(\mu)}{\pi}(\gamma_1^{c\bar{c}}+2c_1)+
 (\frac{\alpha_s(\mu)}{\pi})^2(\gamma_2^{c\bar{c}}+2c_1\gamma_1^{c\bar{c}}+2c_2-5c_1^2)]~.
  \label{equ:MSgamma}
\end{eqnarray}
The $\gamma_1^{c\bar{c}}$, $\gamma_2^{c\bar{c}}$, $\hat{\Gamma}_{0}^{c\bar{c}}$, $\hat{\Gamma}_{1}^{c\bar{c}}$, and $\hat{\Gamma}_{2}^{c\bar{c}}$ conventions are similar to those used in \cite{Bernreuther:2018ynm}. Here, $\gamma_1^{c\bar{c}}=C_F\, \delta_1 \pi$.  In eq.(\ref{equ:MSgamma}), it can be observed that the large log term $\ln (m_h^2/m_c^2)$ that appears in the $\alpha_s$ order correction is analytically reduced by the log term, $\ln (\mu^2/m_c^2)$, in the $\overline{\text{MS}}$ Yukawa coupling if $\mu \sim m_h$.  At the NNLO QCD,
the corresponding numerical  results  are listed in table \ref{tab:ms}  for three different renormalization scales. In fig.\ref{fig:sta}, we show the dependence of $\overline{\Gamma}_{I}^{c\bar c}$ (I=LO, NLO, NNLO) on $\mu$.
Here, we use $\overline{\Gamma}_{LO}^{c\bar c}$,  $\overline{\Gamma}_{NLO}^{c\bar c}$ and
$\overline{\Gamma}_{NNLO}^{c\bar c}$ to respectively
denote the LO, NLO, and NNLO QCD corrections in terms of the $\overline{\text{MS}}$ Yukawa coupling.
The $\mu$ dependence is significantly decreased at the NNLO QCD.
We also show the differential distribution of
the $c\bar{c}$ invariant mass in Fig.\ref{fig:mcc}, where sub-process $h\to c\bar c c\bar c$ is not included to avoid confusion.
In the small $M_{c\bar{c}}$ region,  the NNLO QCD  contributions are larger  than NLO QCD corrections because the  soft  and/or collinear
gluon radiation becomes significant.
\begin{table}[!ht]
  \centering \caption{Inclusive decay width of $h\to c\bar{c}$  in terms of $\overline{\text{{MS}}}$ Yukawa coupling.}\label{tab:ms}
\begin{tabular}{|c|c|c|c|}\hline\hline
          ~~~      & $\mu~=~m_h/2$       ~~~     &$\mu~=~m_h$      ~~~    &$\mu~=~2m_h$    \\[1pt]\hline
        $\overline{\Gamma}_{LO}^{c\bar{c}}$[MeV]    ~~~       &0.1033          ~~~    &0.0916           ~~~   & 0.0823   \\\hline
        $\overline{\Gamma}_{NLO}^{c\bar{c}}$[MeV]    ~~~       &0.1153          ~~~    & 0.1103         ~~~   & 0.1049  \\\hline
        $\overline{\Gamma}_{NNLO}^{c\bar{c}}$[MeV]    ~~~       &0.1162         ~~~    & 0.1148        ~~~   & 0.1125  \\\hline \hline
\end{tabular}
\end{table}

To obtain the precise result for $h\to c\bar{c}$, the NLO  EW corrections should be included along with the NNLO QCD corrections.
The corresponding decay width at the NLO  EW can be written as follows
 \begin{align}
 \Gamma_{EW}^{c\bar c}\, =\, \overline{\Gamma}_{LO}^{c\bar c} ~\Big[  \mathcal{O}_{QED}^{c\bar c}\, + \, \mathcal{O}_{Weak}^{c\bar c} \Big] ~.
\label{equ:EW}
 \end{align}
The NLO QED correction, $\mathcal{O}_{QED}^{c\bar c}$  has a similar form as that of the NLO QCD, i.e.,
 \begin{align}
\mathcal{O}_{QED}^{c\bar c}\, =\, &\alpha\, Q^2_c\, \delta_1 ~,
\label{equ:OEW}
 \end{align}
where  $\alpha$ is the QED coupling, and we choose $\alpha(m_Z)=1/127.934$ \cite{Tanabashi:2018oca}.  $Q_c$ denotes the electric charge of the charm quark.
The coefficient $\delta_1$ at the NLO QED is similar to that at the NLO QCD.
The large logarithm $\ln (m_h^2/m_c^2)$ in $\delta_1$ at the NLO QED can be absorbed in the running  quark mass as in the QCD corrections.
For the NLO weak correction $ \mathcal{O}_{Weak}^{c\bar c}$, there are several studies on this topic \cite{Dabelstein:1991ky,Fleischer:1980ub}. In this study, we adopt the approximation of \cite{Kniehl:1991ze}.
 \begin{align}
\mathcal{O}_{Weak}^{c\bar c}\, =\, & \frac{\alpha}{16 m_W^2 \pi}\big \{k_c m_t^2+m^2_W[-5+\frac{3}{s^2_W}\ln  c_W^2]-8m_Z^2(6v^2_{Zc\bar c}-a^2_{Zc\bar{c}})\big \}~,
\label{equ:OEW}
 \end{align}
where $v_{Zc\bar{c}}=I^c_3/2-Q_c s_W^2$ and $a_{Zc\bar{c}}=I^c_3/2$ with $I_3^c$ denoting the third component of the electroweak isospin  of the charm quark, and
 the coefficient $k_{c}=1$. $s_W=\sin\theta_W$ and $c_W=\cos\theta_W$ with the $\theta_W$  the weak angle, and we choose $m_W=80.358$ GeV, and $s_W^2=0.2233$ \cite{Tanabashi:2018oca} in our numerical calculations. Finally, we obtain the total decay width of the $h\to c\bar{c}$ including the NNLO QCD and NLO
EW corrections as follows
 \begin{align}
\Gamma_{total}^{c\bar c}\, =\, \overline{\Gamma}^{c\bar c}_{NNLO}+ \Gamma_{EW}^{c\bar c}~.
\label{equ:total}
 \end{align}
To combine the precise theoretical predictions and experimental measurements for $h\to f\bar f$ with $f=c$, $b$, we can abstract  some useful information of the
running mass effect in the Yukawa couplings. To acieve this, we list the results for the decay width of $h\to c\bar c$ at the NNLO QCD and NLO EW, and
that of $h\to b\bar b$  together with their ratio in table \ref{tab:ratioMS} for three renormalization scales.
Ratio  ${\Gamma}^{b\bar{b}}_{total}/{\Gamma}^{c\bar{c}}_{total}$ is approximately twenty.
Therefore, it is possible to measure the $h\to c\bar c$ events at the LHC because approximately $1.5\times10^5$ $h\to b\bar b$ events have been observed with data samples corresponding to the integrated luminosity of $79.8~ fb^{-1}$ via the $W h$ and $Z h$ associated production \cite{Sirunyan:2018kst,Aaboud:2018zhk}.  Therefore, approximately $2.8\times 10^4$ $h\to c\bar c$ events will be produced in the similar way but with $300~ fb^{-1}$ integrated luminosity, which make it possible to observe $h\to c\bar{c}$.
 However, the small decay width of $h\to c\bar c$ makes it  advantageous  to search for new physics that
may induce relatively large interactions related to the charm quark,
for example, the ALP  discussed  in the next section.

\begin{table}[!ht]
  \centering \caption{Total decay width ${\Gamma}^{f \bar f}_{total}$  of $h\to f \bar f$
with $f\bar{f}=c\bar{c}, b\bar{b}$ at the NNLO QCD  and NLO EW, together with their ratio.}\label{tab:ratioMS}
\begin{tabular}{|c|c|c|c|}\hline\hline
 ~~~      & $\mu~=~m_h/2$       ~~~     &$\mu~=~m_h$      ~~~    &$\mu~=~2m_h$    \\[1pt]\hline
 ${\Gamma}^{c\bar{c}}_{total}$~[MeV]    ~~~       &0.1165          ~~~    &0.1151           ~~~   &0.1129   \\\hline
 ${\Gamma}^{b\bar{b}}_{total}$~[MeV]    ~~~       &2.4248         ~~~    &2.3990         ~~~   &2.3533  \\\hline
  ${\Gamma}^{b\bar{b}}_{total}/{\Gamma}^{c\bar{c}}_{total}$    ~~~       &20.8137         ~~~    &20.8427          ~~~   & 20.8441  \\\hline \hline
\end{tabular}
\end{table}

\section{ALP production in Higgs decay}\label{Sec:ALP}

We begin by considering a general ALP  with flavor violating couplings to the right-handed up-quarks \cite{Carmona:2021seb}. To describe such a system, the most general effective field theory is given by the following lagrangian
\begin{align}
-{\cal L}&\, =\, \frac{1}{2}(\partial_{\mu}a)(\partial^{\mu}a)-\frac{m_a^2}{2}a^2+\frac{\partial_{\mu}a}{f_a}
[(c_{uR})_{ij}\bar{u}_{Ri}\gamma^{\mu}u_{Rj}+c_{\Phi} \Phi^{\dag}i\overleftrightarrow{D}_{\mu}\Phi]\nonumber\\
&-\frac{a}{f_a}[c_g \frac{g_3^2}{32\pi^2} G^{a}_{\mu \nu} \tilde{G}^{\mu \nu a}
+c_W\frac{g_2^2}{32\pi^2} W_{\mu \nu}^I \tilde{W}^{\mu\nu I}+c_B \frac{g_1^2}{32\pi^2}B_{\mu \nu}\tilde{B}^{\mu\nu}]~,
 \label{equ:La}
\end{align}
where $a$ and $\Phi$ denote the ALP and Higgs field, respectively.
 $g_1$, $g_2$, $g_3$ are the $U(1)_Y$, $SU(2)_L$ and $SU(3)_c$  gauge couplings of the SM, respectively, whereas
$B_{\mu \nu}$, $W_{\mu \nu}^I$, I=1,2,3, and $G^b_{\mu \nu}$, b=1,...8, are their corresponding field-strength tensors.  $\tilde{B}_{\mu\nu}=1/2 \epsilon_{\mu\nu\alpha\beta} B^{\alpha\beta}$, ..., represent the corresponding dual field.
$U_{Ri}$ with $i=1,2,3$ denotes the right-handed SM up-quark of the $i$th generation. $f_a$ can be treated as a free parameter.
$c_{\Phi}$, $c_g$, $c_W$, and $c_B$ are the Wilson coefficients, and $c_{uR}$ is a hermitian matrix. In this model, it is assumed that the ALP is a pseudo Nambu-Goldstone boson of the spontaneous breaking of a global $U(1)$ symmetry, and
 that its couplings to the leptons, SM quark doublets and right-handed down-type quarks vanish.  The operator
$(\partial^{\mu}a/{f_a})  \Phi^{\dag}i\overleftrightarrow{D}_{\mu}\Phi$ can be traded using Higgs field redefinition \cite{Brivio:2017ije}.

\begin{figure}[ht]
   \centering
   \includegraphics[scale=0.8]{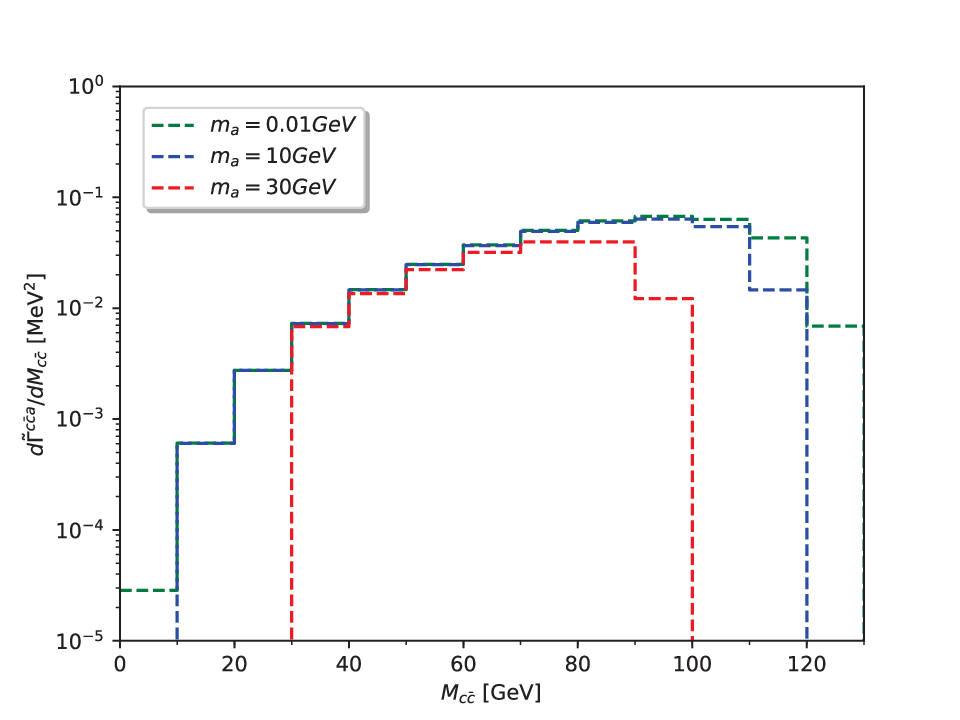}
   \caption{Distribution $d\tilde{\Gamma}^{c\bar{c}a}/dM_{c\bar{c}}$ of the $c\bar{c}$ invariant mass.}
   \label{fig:Mcca}
\end{figure}

\begin{figure}[ht]
   \centering
   \includegraphics[scale=0.5]{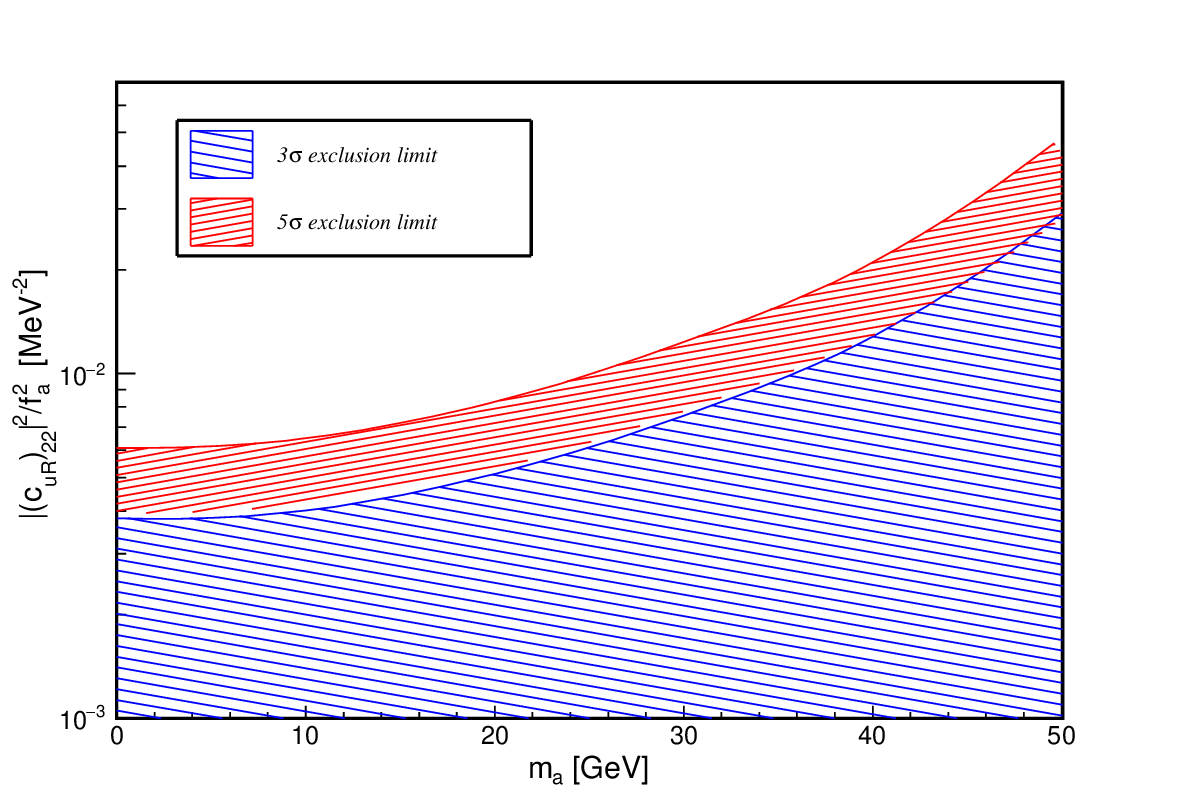}
   \caption{3$\sigma$ exclusion limit (blue) and 5$\sigma$ discovery limit (red) of $h\to c\bar{c}a$ at ILC.}
   \label{fig:ee}
\end{figure}

We then evaluate the ALP  production in association with the charm quark pair in the Higgs boson decay.
The differential decay width of the $h\to c\bar{c} a$ process where $a$ denotes the ALP can be
obtained as follows
\begin{eqnarray}
d\Gamma_{h\to c\bar{c} a} & \, =\,  & \frac{1}{2 m_h}\,  |{\cal M}|^2\,  d{\cal L}ips_3~,
\end{eqnarray}
where ${\cal L}ips_3$ is the Lorentz invariant phase space of the final three particles. The matrix element square can be expressed as follows
\begin{eqnarray}
|{\cal M}|^2 &\, =\,  &\frac{2 y_c^2\,\, |(c_{uR})_{22}|^2}{f_a^2}\,
\Big \{  {m_h}^2({x_c}+{x_{\bar{c}}}-1-\frac{4{m_c}^2}{{m_h}^2}+\frac{{m_a}^2}{{m_h}^2})+  \frac{1}{{m_h}^2} \Big[
\frac{-{m_c}^2{m_h}^2(x_c-x_{\bar{c}})^2}{(1-{x_c}) (1-{x_{\bar{c}}})} \nonumber \\
&&+\, \frac{{m_a}^2}{{m_h}^2} \Big(
\frac{{m_c}^2(4{m_c}^2+{m_h}^2(1-2x_{c}))}{(1-{x_{c}})^2}
+\frac{{m_c}^2(4{m_c}^2+{m_h}^2(1-2x_{\bar{c}}))}{(1-{x_{\bar{c}}})^2} \nonumber \\
&& +\, \frac{2{m_c}^2(4{m_c}^2-{m_h}^2({x_c}+{x_{\bar{c}}}-1))}{(1-{x_c}) (1-{x_{\bar{c}}})} \Big)\Big ]
\Big \} ,
\label{eq:mm}
\end{eqnarray}
where $m_a$ is the ALP mass. $x_c=2E_c/m_h$ and $x_{\bar{c}}=2E_{\bar{c}}/m_h$ with $E_c$($E_{\bar{c}}$), the $c$($\bar c$) quark energy in Higgs rest frame.
In eq.(\ref{eq:mm}), the dependence of the ALP mass appears in the form $\mathcal{O}(m_a/m_h)^0+\mathcal{O}(m_a/m_h)^2$, where $(m_a/m_h)^2$ is a tiny quantity for $m_a\ll m_h$ so that the height of the distribution is not sensitive to the ALP mass.
In the following numerical calculations, we factor out $|(c_{uR})_{22}|^2/f_a^2$ from $\Gamma_{h\to c\bar{c} a} $,  i.e.,
\begin{equation}
\Gamma_{h\to c\bar{c} a}\,  =\, \frac{|(c_{uR})_{22}|^2}{f_a^2}\, \tilde{\Gamma}_{h\to c\bar c a}~.
\end{equation}
The experiment searches on the ALP parameter spaces that is $m_a$ and $f_a$ allow access to several orders of magnitude. The joint limit of these two parameters is given experimentally.
Direct searches for the ALP and calculations of their effect on the cooling of stars and on the supernova SN1987A
impose $f_a\gtrsim 4\times 10^8$ GeV \cite{Raffelt:2006cw}.
The thermally produced ALP DM is allowed in sizable parts of the parameter $m_a\gtrsim 154$ eV \cite{Cadamuro:2010cz}.
 In our analysis, $f_a$ is treated as a free parameter that relaxes the restriction of $m_a$. We choose $m_a=0.01, ~ 10, ~30$ GeV.
The results for the differential distribution with respect to $M_{c\bar c}$  are shown in fig.\ref{fig:Mcca}.
Because there is no exact measurement for $h\to c\bar c$, to obtain some constrain on the parameters, we set the condition
\begin{equation}
\Gamma_{h\to c\bar{c} a}\,  \leq 1\%\, {\Gamma}^{c\bar{c}}_{total}\; , ~~\text{and} ~~10\%\,  {\Gamma}^{c\bar{c}}_{total}.
\label{ALP1}
\end{equation}
 \begin{figure}[ht]
   \centering
   \includegraphics[scale=0.5]{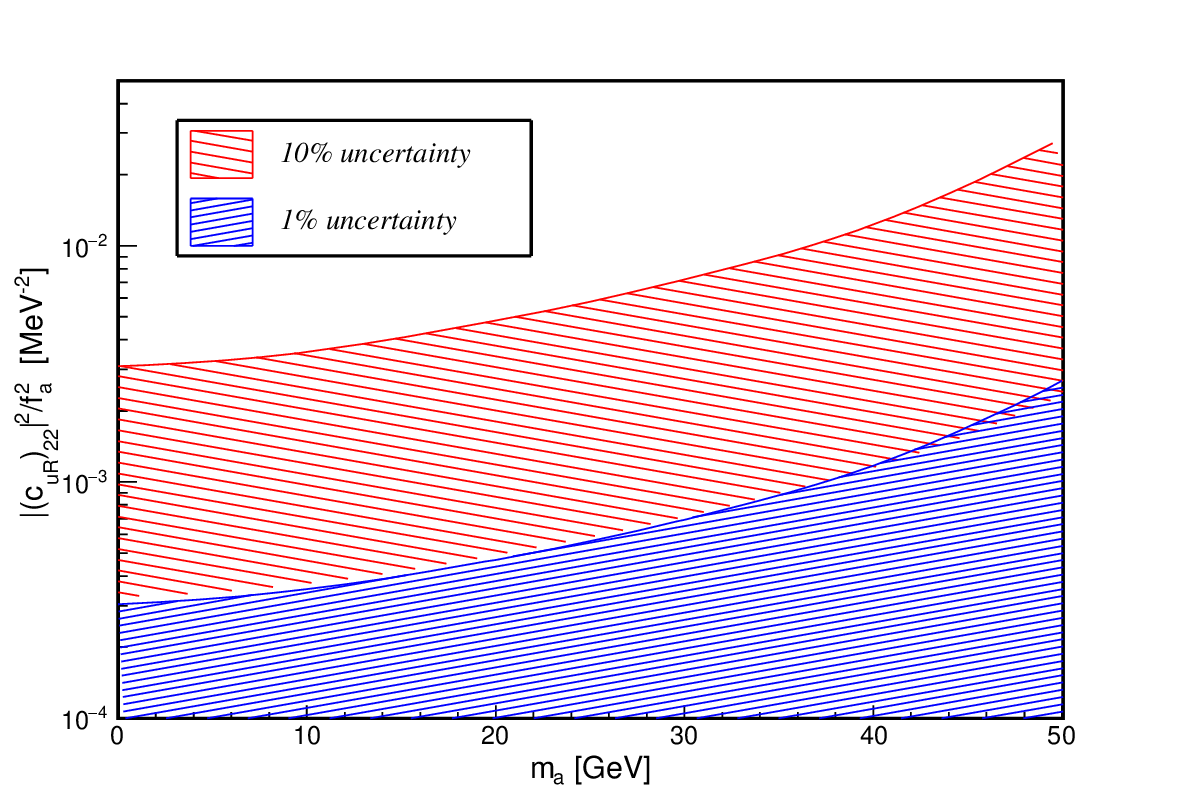}
   \caption{Constraints on $|(c_{uR})_{22}|^2/f_a^2$  as a function of $m_a$, resulting from the conditions $\Gamma_{h\to c\bar{c} a}\,  \leq 10\%\, {\Gamma}^{c\bar{c}}_{total}$ and $\Gamma_{h\to c\bar{c} a}\,  \leq 1\%\, {\Gamma}^{c\bar{c}}_{total}$, with the allowed regions marked by red and blue, respectively.}
   \label{fig:DWa}
\end{figure}
The constraints for $|(c_{uR})_{22}|^2/f_a^2$ with respect to $m_a$  are shown in fig.\ref{fig:DWa}. A loose constrain for the larger ALP mass can be found. It is also interesting to search for the ALP via the $h\to c\bar{c}a$ process in the future electron-positron collider, e.g. the Circular Electron-Positron Collider (CEPC) or International Linear Collider (ILC). By comparing the distribution of the $h\to c\bar{c}$ background and $h\to c\bar{c}a$ signal in Fig.\ref{fig:mcc} and Fig.\ref{fig:Mcca}, it can be observed that for the background, the distribution tends to be close to the Higgs mass, whereas the signal process does not. Therefore, for a rough estimation, we require that the invariant mass of $c$, $\bar{c}$ be less than 110 GeV. Furthermore, we consider the $e^+ e^- \to Zh$ process with $Z\to e^+ e^-, \mu^+ \mu^-$ and $h\to c\bar{c}$ or $c\bar{c}a$ at $\sqrt{s}=250$ GeV. The 3$\sigma$ exclusion limit and 5$\sigma$ discovery limit with the integrated luminosity of 250 $fb^{-1}$ are shown in fig.\ref{fig:ee}.
When the high precision measurement of $h\to c\bar c$ is available, the realistic constrain for the ALP
production in $h\to c\bar c$ can be obtained.

\section{Summary}\label{Sec:Sum}
Measurements on the $h\to c\bar c$  with high statistics at the LHC are still in progress. More theoretical studies on this process are still necessary and significant. Moreover, after the discovery of the SM Higgs, the search for new (pseudo) scalar boson beyond the SM, e.g., ALP, has become a significant topic in particle physics.
In this paper, we present the results of the decay widths for $h\to c\bar{c}$ at the NNLO QCD and NLO EW corrections.
For the flavor-singlet contributions where the Higgs boson coupled to the bottom and top quark appeared at an order of $\alpha_s^2$, we provide the exact results, and
find that the exact result of the top quark triangle is very close to the approximate result calculated to the large top quark mass expansion.
The results for the Yukawa coupling  defined in the $\overline{\text{MS}}$ scheme
is more reliable and the large logarithmic effect related to the ratio of the charm-Higgs mass is reduced.
Finally, we evaluate the ALP associate production with the charm quark pair, and  the constrain
for the related parameters is estimated by assuming a condition of eq.(\ref{ALP1}).
At the upgraded LHC, an increasing number of events  on the Higgs boson decay will be accumulated, so that precise studies on $h\to c\bar c$ and the search for new particles in the Higgs decay
become possible.

\section*{Acknowledgement}
The authors thank Profs. W. Bernreuther, H. F. Li and Dr. L. Chen for their helpful discussions, and also thank Prof.  F. Tramontano for providing us the code to calculate the virtual flavor-singlet triangle loop. This work is supported in part by national science foundation of China under the grant
Nos. 11875179, 11775130 and 11635009.

\end{document}